\documentclass[%
 reprint,
superscriptaddress,
amsmath,amssymb,
aps,
pra,
]{revtex4-2}
\usepackage{caption}
\usepackage{subcaption}
\usepackage{graphicx}
\usepackage{xcolor}
\usepackage{graphicx}
\usepackage{dcolumn}
\usepackage{bm}
\usepackage{upgreek}
\usepackage{lineno}

\begin{document}

\preprint{APS/123-QED}

\title{Resonance-free Fabry-P{\'e}rot cavity via unrestricted orbital-angular-momentum ladder-up}

\author{Shaghayegh Yaraghi$^{1}$, Oussama Mhibik$^{1}$, Murat Yessenov$^{1,2}$, J. Keith Miller$^{1}$, Midya Parto$^{1}$, Eric G. Johnson$^{1}$, Ayman F. Abouraddy$^{1}$, and Ivan Divliansky}
\email{Corresponding author: ibd1@creol.ucf.edu}
\affiliation{CREOL, The College of Optics \& Photonics, University of Central Florida, Orlando, Florida 32816, USA\\
$^{2}$Harvard John A. Paulson School of Engineering and Applied Sciences, Harvard University, Cambridge, MA, USA}

\begin{abstract}
Introducing elements into an optical cavity that modify the transverse spatial field structure can also impact the cavity spectral response. In particular, an intra-cavity spatial mode-converter is expected to induce modal runaway: unrestricted ladder-up in the modal order, concomitantly thwarting coherent field interference, thereby altogether suppressing the resonant response -- a phenomenon that has yet to be observed in an optical cavity. Here we show that a single intra-cavity holographic phase mask placed in a compact free-standing planar Fabry-P{\'e}rot cavity renders the cavity spectral response resonance-free. By acting as a mode-converter on a basis of Laguerre-Gaussian (LG) modes, an incident broadband fundamental Gaussian mode exits the cavity in the form of a superposition of a large number of collinearly propagating broadband LG modes of fixed parity whose spectra coincide with that of the input. Crucially, the resonance-free spectral response is maintained while changing the cavity length by $\sim350\%$, raising the prospect of stable resonant optical sensors whose performance is impervious to length perturbations.
\end{abstract}

\maketitle


Optical resonators are a key building block in optics and photonics, which have enabled the optical revolution that started with the invention of the laser \cite{SalehBook19}. Resonant field buildup resulting from intra-cavity light recycling is useful across a broad swath of applications: enhancing linear effects (e.g. coherent perfect absorption) \cite{Wan11S,Pye17OL,Baranov17NRM,Villinger21AOM}, nonlinear effects (e.g., harmonic generation and nonlinear wavelength conversion) \cite{Turner08OE,Hashemi09PRA,Lin16Optica,Ji17Optica,Koshelev20Science}, sensing \cite{Armani07Science}, optical combs \cite{Pfeifle14NP} and clocks \cite{Diddams01Science}, in quantum optics \cite{Haroche2006book}, and even in gravitational wave detection \cite{Wise04CQG}. Nevertheless, light recycling is achieved coherently only over a narrow linewidth at discrete resonant wavelengths separated by the free spectral range (FSR). Moreover, further enhancing the field buildup can be harnessed only over an ever-shrinking linewidth \cite{SalehBook19}. This fundamental complementarity between field buildup and resonant linewidth has spurred on efforts towards the realization of linewidth broadening without sacrificing the cavity quality factor. Examples along these lines include white-light cavities \cite{Wicht97OC,Rinkleff05PS} that incorporate an active medium endowed with anomalous dispersion (whether atomic species \cite{Pati07PRL,Wu08PRA} or nonlinear effects \cite{Yum13JLT}), omni-resonant cavities \cite{Shabahang17SR} in which angular dispersion is introduced into the externally incident field to couple a broad continuous spectrum into the cavity \cite{Shabahang19OL,Shiri20OL,Shiri20APLP}, among other efforts \cite{Savchenkov06OL,Kotlicki14OL,Strekalov15OL,Lin15OL}.

A different research enterprise has recently focused on modifying the transverse spatial structure of a laser field through introducing intra-cavity optical elements \cite{Forbes24NPR}; for example, amplitude stops or apertures \cite{Oron01PO}, spatial light modulators \cite{Ngcobo13NC,ZhouSR18}, phase plates \cite{Naidoo16NP}, or metasurfaces \cite{Maguid18ACSP,Sroor20NP,Piccardo22NP}. These efforts have yielded lasers emitting prescribed orbital angular momentum (OAM) modes or other spatial field structures \cite{Forbes21NP}. Only recently has it  been recognized that intra-cavity modification of the \textit{spatial} structure of the cavity field can exert a profound impact on its \textit{spectral} structure \cite{Ginis23NC}, thereby opening the possibility of spectrally broadening of the resonant bandwidth through spatial structuring of light. However, the long cavity lengths typically utilized for producing structured laser fields unfortunately mask any such effects by rendering the free spectral range (FSR) too small to be readily resolved, thus precluding the investigation of coupling between the spatial and spectral degrees-of-freedom.

This limitation can be lifted by exploiting on-chip waveguide-based cavities, which facilitates spectrally resolved measurements by reducing the cavity length (e.g., length $\sim150$~$\mu$m with FSR $\sim1.5$~nm at a wavelength of 1.5~$\mu$m in \cite{Ginis23NC}). Intriguingly, the impact of an intra-cavity mode-converter on the cavity spectral response is sensitive to the modal basis selected. In the case of a waveguide geometry that supports Hermite-Gaussian-like (HG-like) modes in a Cartesian coordinate system, a single intra-cavity mode-converter toggles the cavity field between two waveguide modes after each cavity roundtrip, effectively doubling the cavity length and thus halving its FSR \cite{Ginis23NC}. Further modification of the \textit{spectral} structure in such a waveguide-based cavity requires introducing a \textit{sequence} of spatial mode-converters -- each responsible for converting between a particular pair of HG modes. In this case, a restricted ladder-up effect of the modal order can be produced (limited by the number of individual mode-converters implemented), which can be used to establish a folded back-and-forth light-recycling path with broadband operation as realized in \cite{Chang17arxiv}. However, a cavity structure was not established in \cite{Chang17arxiv}, so that broadband operation is guaranteed in absence of the mode-converters, albeit with only a single pass through the waveguide. To rely on a \textit{single} mode-converter to achieve this effect therefore requires changing the modal basis.

\begin{figure*}[t!]
\centering
\includegraphics[width=17.6cm]{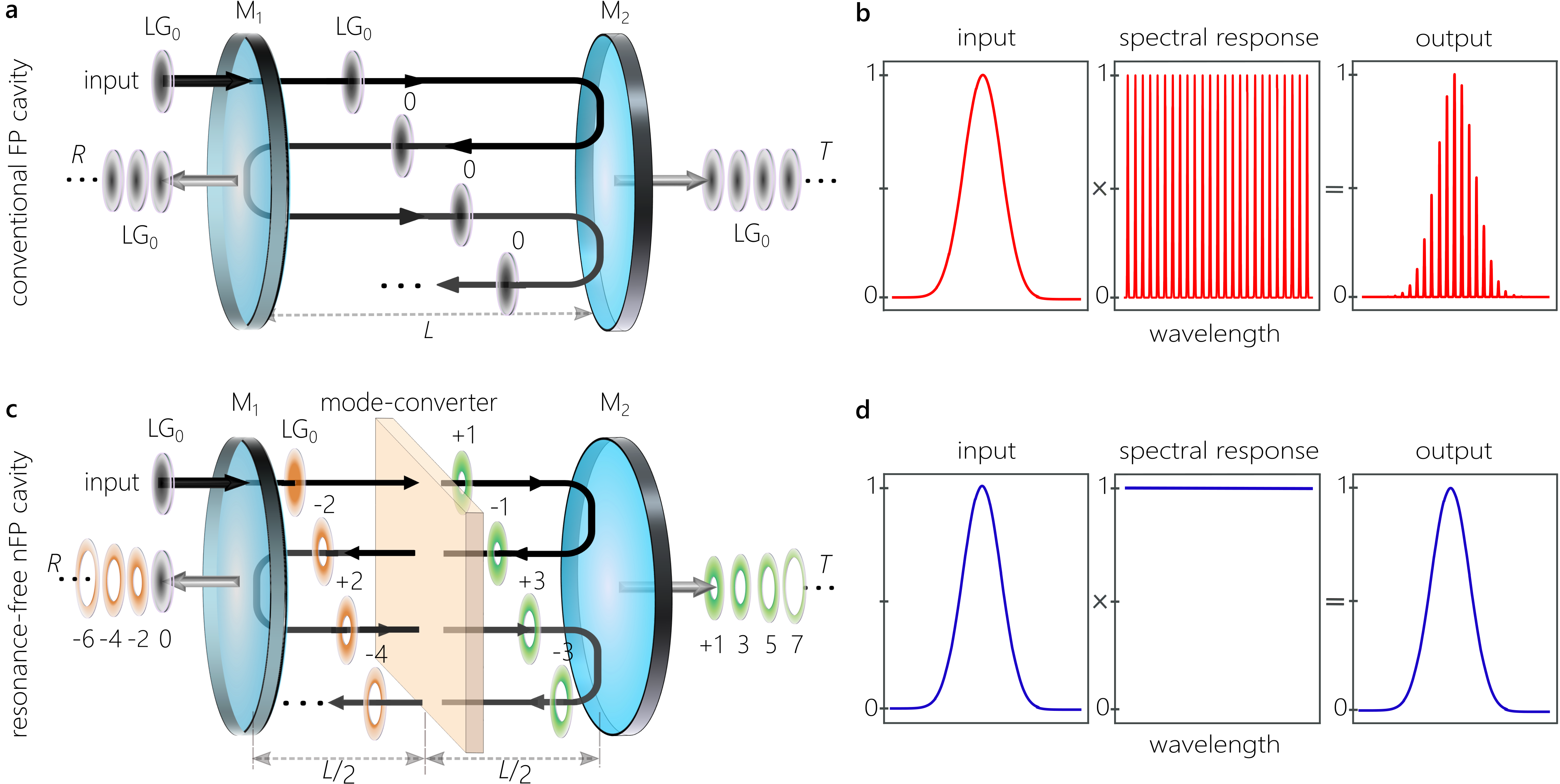}
\caption{\small{\textbf{Conventional FP and resonance-free nFP cavities.} (a) Conventional FP cavity consisting of two planar mirrors M$_{1}$ and M$_{2}$ illuminated by a Gaussian beam LG$_0$. $R$: Reflection; $T$: transmission; $L$: cavity length. (b) The spectral response of the FP cavity is dominated by the usual resonance peaks. The output spectrum is the product of the input spectrum and the cavity spectral response. (c) A resonance-free nFP cavity consisting of mirrors M$_{1}$ and M$_{2}$ and containing an HPM as a mode-converter in the LG modal basis. The OAM order of the field increases by 2 after each roundtrip. The orthogonality of the LG modes thwarts field interference. In (a) and (b) we list only the modal order $m$ of the LG$_{m}$ modes in the subsequent roundtrips for clarity. (d) The spectral response of the nFP cavity is flat and resonance-free. Consequently, the output spectrum coincides with the input spectrum.}}
\label{Fig:concept}
\end{figure*}

It is therefore intuitively expected that an intra-cavity mode-converter can, in principle, result in \textit{unrestricted} modal ladder-up (or modal run-away), so that a single spatial mode externally incident on this cavity produces an output in the form of a superposition of a large number of collinearly propagating orthogonal modes. Furthermore, accompanying this complex spatial field response is a dramatic change in the spectral structure. Rather than the discrete spectral resonances, the orthogonality of the cavity transverse modes thwarts intra-cavity interference, thereby suppressing the resonant response. Such a cavity is expected to display a broadband resonance-free spectral response that is altogether independent of the cavity length

Nevertheless, the use of a single intra-cavity element to enable modal ladder-up and a resonance-free spectral response has eluded observation to date in an optical cavity. Despite its ceonceptual intuitiveness, this goal represents a substantial technical challenge. First, the cavity needs to remain compact after introducing the intra-cavity elements to facilitate spectrally resolving the resonances. Second, as mentioned above, waveguide HG-like modes are not conducive to modal ladder-up with a single intra-cavity element, and modes carrying orbital angular momentum (OAM) are preferred instead because a single phase element can implement a modal transformation in this basis for all mode orders. Third, high-efficiency modal conversion is required to eliminate any remnant resonant spectral response without deleteriously impacting the cavity finesse. Fourth, the orthogonality of the modes produced by the mode-converter is required to suppress intra-cavity interference of the cavity field. Fifth, large-area modes are necessary to reduce diffraction over the entire resonant path length in the cavity, to further maintain mode-orthogonality in all the roundtrips. These requirements place stringent constraints on the cavity design and construction to achieve a resonance-free spectral response. 

\begin{figure*}[t!]
\centering
\includegraphics[width=17.6cm]{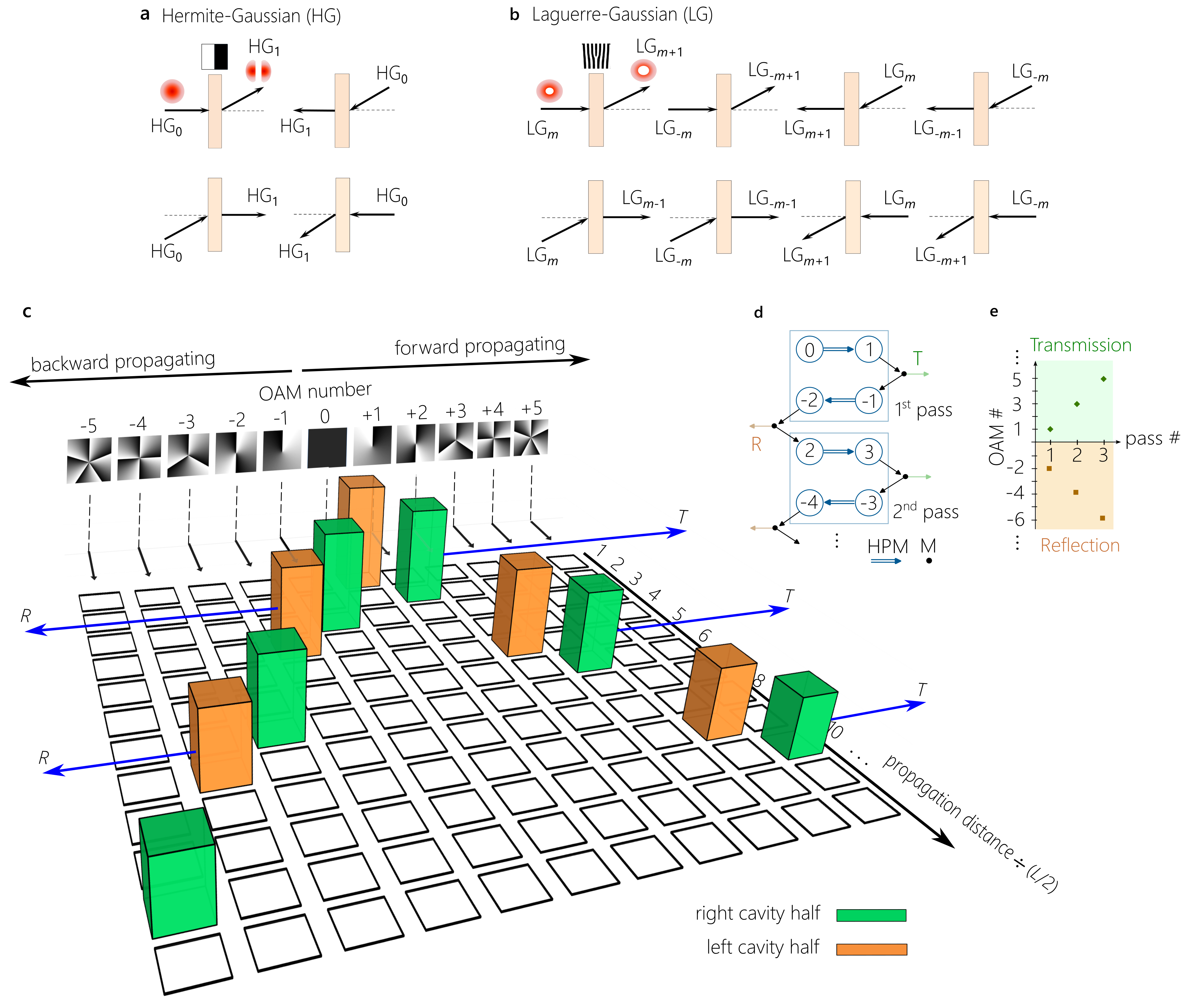}
\caption{\small{\textbf{Modal conversion in an nFP cavity.} (a) Modal conversion in the HG basis with a $\pi$-step phase plate, which toggles the HG$_{0}$ and HG$_{1}$ modes. (b) Configurations for modal conversion in the LG basis with an HPM (experimental confirmation for all 8 configurations is provided in Supplementary Section~1 and Supplementary Figure~1). (c) The evolution of the modal structure in an nFP cavity comprising the HPM when LG$_{0}$ is externally incident from the left (Fig.~\ref{Fig:concept}c). The phase distributions displayed at the top correspond to the LG mode in each column. The left and right halves of the matrix correspond to forward and backward propagating fields (with respect to the incident field), respectively. Progression along the propagation axis corresponds to the successive roundtrips, and the orange and green columns correspond to even and odd LG modes, respectively. The height of the columns gradually drops with the amplitude factor associated with each roundtrip. The reflected ($R$) and transmitted ($T$) modes exiting the nFP cavity are identified with arrows. (d) A directed graph description of the modal-order (the encircled numbers) evolution, and (e) an illustration of the unrestricted modal-order ladder up with propagation of an incident LG$_{0}$ mode in the nFP cavity.}}
\label{Fig:ModalEvolution}
\end{figure*}

Here we show that introducing a single, thin, intra-cavity phase element into a planar Fabry-P{\'e}rot (FP) cavity can radically change the cavity spectral response to yield a broadband resonance-free cavity. Upon incorporating the phase element, the conventional discrete resonant spectral structure is entirely eliminated, and the cavity spectral response is immediately rendered flat. By utilizing a free-standing yet compact FP cavity, we satisfy multiple desiderata simultaneously. First, any modal basis can be utilized, and we make use here of OAM-carrying Laguerre-Gaussian (LG) modes. The intra-cavity element is a holographic phase mask (HPM) written in a thin volume Bragg grating in transmission mode capable of changing the topological charge (OAM) of an LG mode traversing it by $\pm1$ with high conversion efficiency ($>85\%$ for all modal orders), so that the cavity field increases the  OAM by 2 after each roundtrip. Second, the FP cavity accommodates a large transverse beam profile whose associated Rayleigh range significantly exceeds the distance corresponding to the cavity photon lifetime (the product of the cavity finesse and roundtrip length), and is simultaneously compact to yield a large FSR. All the forward and backward propagating field components are mutually orthogonal, which thwarts coherent field interference and frustrates the emergence of resonances in the transmitted spectrum. We are therefore able to carefully correlate the elimination of the resonant spectral structure with the modification of the transverse spatial field structure. We have unambiguously verified our predictions experimentally using a variety of measurement approaches. We first verify the performance of the free-standing HPM as a mode-converter carrying out the OAM transformation LG$_{0,m}$ to LG$_{0,m+1}$. After introducing the HPM into the cavity, we resolve the modal content produced by an incident broadband fundamental mode LG$_{00}$ via an LG mode-sorter that reveals modes extending up to $m=19$ collinearly produced from the cavity. We next measure the spectral response of the cavity in the absence and presence of the HPM: whereas the former displays the usual discrete resonant spectral lines, the latter has a flat spectral response devoid of any resonances, so that the output spectrum coincides with the incident spectrum. We have confirmed this resonance-free behavior while changing the cavity length by $\approx350\%$ (from 1.3~mm to 4.8~mm). Whereas the FSR of the cavity in absence of the HPM changes accordingly (from 274~pm to 100~pm) over this span of cavity lengths, the spectral response in the presence of the HPM remains invariant. These results may have profound implications for lasing in active micro-cavities in which the spatial field structure is modified, while offering opportunities for stable resonant optical sensors that are impervious to cavity-length perturbations, in addition to providing a platform for exploring the non-Hermitian dynamics of exceptional points \cite{Elganainy18NPhys} along the dual synthetic dimensions of modal order and wavelength.

\noindent\textbf{Conventional FP cavity versus resonance-free FP cavity.} A conventional FP cavity confines light between two mirrors, thereby leading to field-buildup in a standing wave at specific resonant frequencies (Fig.~\ref{Fig:concept}a,b) \cite{SalehBook19,Collins60PRL}. The transverse spatial structure of the resonant field is maintained after each cavity roundtrip, except for an overall amplitude factor (Fig.~\ref{Fig:concept}a). The transmission spectrum of an FP cavity comprises sharp peaks (Fig.~\ref{Fig:concept}b) separated by the free spectral range (FSR). For example, when a Gaussian beam is externally incident on this FP cavity, the beam reflects back and forth between the two mirrors, and multipass interference yields a Gaussian beam emerging on resonance, and the spectrum of the Gaussian beam is modulated by the cavity spectral response (Fig.~\ref{Fig:concept}b). We assume that the transverse size of the field is sufficiently large that diffractive effects can be neglected over the entire path length of the field within the cavity. In this case the spatial and spectral responses of the cavity here are \textit{separable}.

Our goal here is to construct a non-standard Fabry-P{\'e}rot (nFP) cavity that is \textit{resonance-free}; i.e., one in which the formation of standing waves is suppressed (Fig.~\ref{Fig:concept}c) to eliminate the resonant response (Fig.~\ref{Fig:concept}d), without changing the number of roundtrips in the cavity. This novel behavior is achieved by introducing a single intracavity purely phase element to act as a spatial mode-converter that increases the modal index $m\rightarrow m+1$ (for all $m$) of a spatial mode belonging to an orthogonal basis, such that $m\rightarrow m+2$ upon each cavity roundtrip (Fig.~\ref{Fig:concept}c). We select here the set of LG$_{0m}$ modes with radial index 0 and azimuthal (OAM) index $m$, and we contract the notation of these pure OAM modes to LG$_{m}$. The mode-converter we utilize here is a holographic phase mask (HPM) that comprises an appropriate phase distribution imprinted within a volume Bragg grating to increase the OAM index by 1 after each pass (Methods).

Several requirements must be satisfied by the HPM for the nFP cavity to exhibit resonance-free behavior. First, high diffraction efficiency (modal conversion) is required; finite modal conversion results in a remnant conventional resonant field component superimposed on the target resonant-free spectral response. Second, a thin component is desirable to minimize the cavity length and yield a large FSR to render the resonance suppression spectrally unambiguous. Third, a small diffraction angle between the incident and diffracted fields is beneficial in optimizing the angular and spectral acceptance of the HPM, and to further ensure a minimal cavity length \cite{SeGall15OE,Kogelnik69BSTJ}, while still eliminating non-diffracted light. Fourth, a large HPM area is desirable to accommodate a large beam profile (and, therefore, a large Rayleigh range). All of these requirements are satisfied by the HPM that has a device thickness of 1~mm, an active area of diameter 3~mm (in a volume Bragg grating of area $12\times12$~mm$^{2}$), and high diffraction efficiency for all modes (as confirmed below).

The intracavity HPM modifies the phase front upon each pass. Traversing from left to right  yields the transformation LG$_{m}\rightarrow$LG$_{m+1}$, and from right to left LG$_{m}\rightarrow$LG$_{m-1}$ \cite{Sroor20NP}. The orthogonality of the LG$_{m}$ modes disrupts constructive multipass interference, yielding a broadband and flat spectral transmission response (Fig.~\ref{Fig:concept}d), devoid of the sharp resonant peaks typical of a conventional FP cavity. Therefore, when a Gaussian beam is incident externally on the nFP cavity, the spatial profile of the transmitted beam comprises a superposition of odd-order LG modes ($m\geq1$), and the spectrum of each mode in the output field coincides with the input spectrum.

\begin{figure}[t!]
\centering
\includegraphics[width=8.6cm]{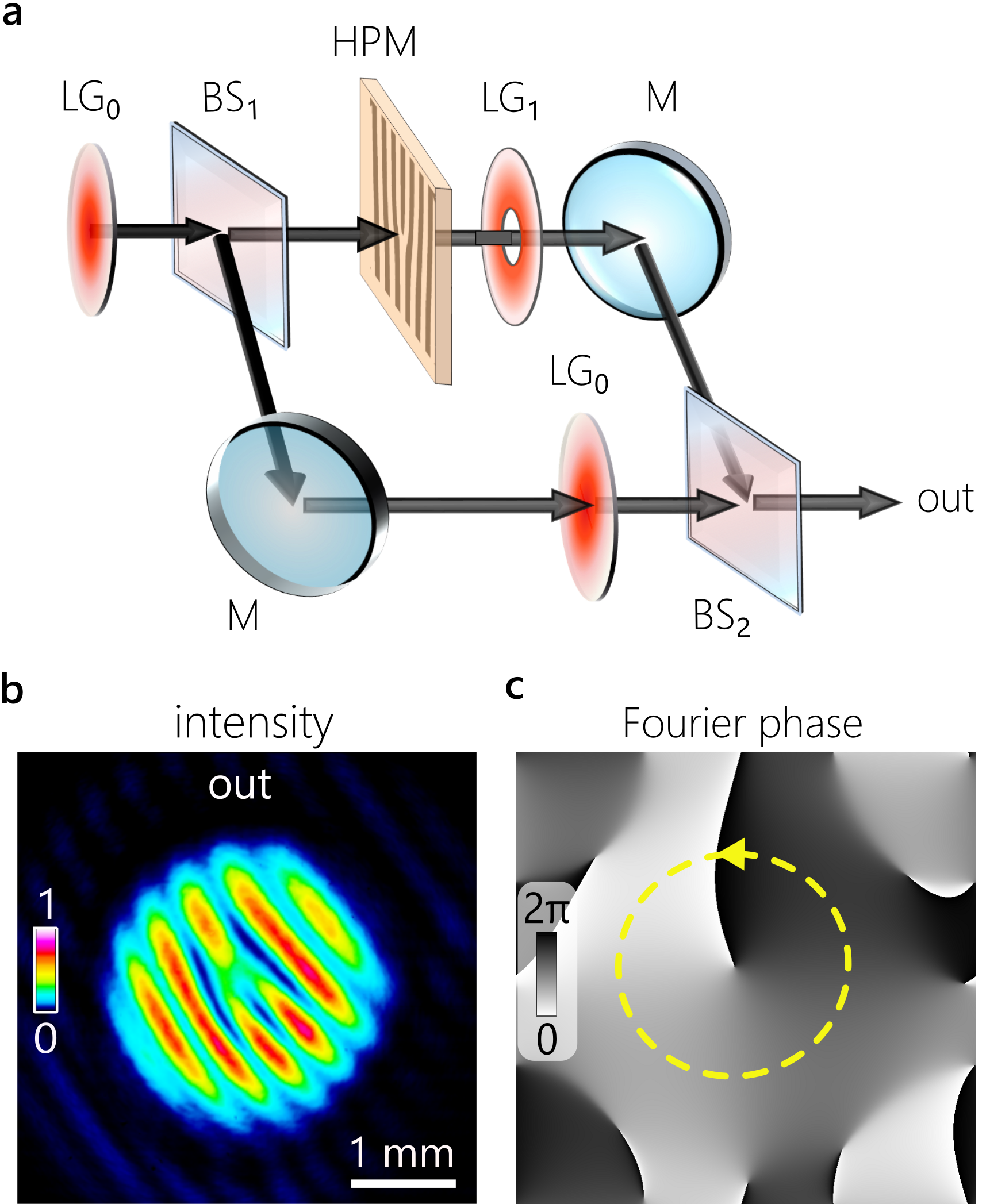}
\caption{\small{\textbf{Phase characterization of the HPM.} (a) A two-path interferometer combines LG$_{0}$ with LG$_{1}$ diffracted by the HPM. BS: Beam splitter; M: mirror. (b) Interference pattern between LG$_{0}$ and LG$_{1}$ at the interferometer output port. (c) Extracted phase map from the interference data in (b) via off-axis holography.}}
\label{Fig:HPMphase}
\end{figure}

\noindent\textbf{Mode conversion in a resonance-free nFP cavity.} We first highlight a key feature of OAM modes in the LG basis in contrast to other modal bases such as that of HG modes. To convert HG$_{0}$ to HG$_{1}$ in one transverse dimension, a $\pi$-phase step is required (Fig.~\ref{Fig:ModalEvolution}a). Traversing from left to right results in HG$_{0}\rightarrow$HG$_{1}$, and from right to left HG$_{1}\rightarrow$HG$_{0}$. Note that reflection from a mirror does \textit{not} impact the modal order of an HG mode. By introducing such a phase pattern in an FP cavity, the field toggles between HG$_{0}$ and HG$_{1}$, no modal ladder-up or run-away occurs, and the resonant spectral structure survives, except for a potential modification of the FSR due to a change in the effective cavity length, as demonstrated recently in \cite{Ginis23NC}. Accessing higher-order HG modes requires utilizing a sequence of individualized mode-converters for each conversion HG$_{m}\rightarrow$HG$_{m+1}$, which was realized in \cite{Chang17arxiv} (up to $m=6$) to fold the propagation path in a multimode waveguide and increase the propagation distance. No cavity was established in \cite{Chang17arxiv}, and thus no resonant response follows when the mode-converters are removed.

A feature common to the work reported in \cite{Ginis23NC} and \cite{Chang17arxiv} is field confinement to an on-chip multimode waveguide that supports HG modes rather than LG modes. Relying instead on LG modes opens the new possibilities illustrated in Fig.~\ref{Fig:ModalEvolution}b. Normal incidence from the left produces a diffracted field on the right in which the OAM index increases by 1 (LG$_{m}\rightarrow$LG$_{m+1}$), while retracing the path from the right to the left decreases the OAM index by 1 (LG$_{m}\rightarrow$LG$_{m-1}$), for all values of $m$; see Supplementary Figure~1 for experimental verification of all the configurations in Fig.~\ref{Fig:ModalEvolution}b. Coupled with the change in the OAM sign upon reflection from a mirror $m\rightarrow-m$, this mode-converter leads to a ladder-up effect in which the modulus of the OAM order progressively increases with each roundtrip (in principle without bound), rather than the modal toggling in the HG basis. Critically, maintaining the orthogonality of these modes further requires complete propagation-collinearity and minimal diffraction over multiple cavity roundtrips.

The evolution of the cavity field is illustrated in the distribution matrix in Fig.~\ref{Fig:ModalEvolution}c. The fields on the right and left sides of the HPM in the nFP cavity (Fig.~\ref{Fig:concept}c) are represented in Fig.~\ref{Fig:ModalEvolution}c by green and orange columns, respectively. A fundamental Gaussian beam LG$_{0}$ entering the cavity diffracts at the HPM to LG$_{1}$, a percentage of which exits the nFP at M$_{2}$. Reflection from M$_{2}$ produces LG$_{-1}$ propagating back to the HPM, whereupon it is converted to LG$_{-2}$, and then to LG$_{2}$ upon reflection from M$_{1}$. Each roundtrip therefore increases the modal index by 2, thereby generating a sequence of LG modes with increasing order, with the forward and backward propagating modes having opposite OAM signs. As shown in Fig.~\ref{Fig:concept}c, only odd LG modes with positive orders are transmitted through the nFP cavity at M$_{2}$, while negatively signed even LG modes exit from M$_{1}$. The effective number of mode conversions is limited by the amplitude reduction upon each roundtrip (i.e., by the cavity quality factor), which is determined by the HPM efficiency, the inherent divergence of the converted LG modes, and the intrinsic cavity roundtrip losses. The modal evolution is further elucidated by the directed graph in Fig.~\ref{Fig:ModalEvolution}d that depicts the ordered sequence of modal conversion over two cavity roundtrips. The plot in Fig.~\ref{Fig:ModalEvolution}e of the modal order with roundtrip number then illustrates the unrestricted modal-order ladder-up that is expected in the nFP cavity.

\noindent\textbf{Modal characterization of the HPM.} The fabricated HPM consists of a volume Bragg grating superimposed with a first-order LG-mode phase profile (Methods). Two distinct measurements verify that the intended HPM functionality is realized: (1) verifying the transverse phase profile imparted by the HPM and the orthogonality of the cavity LG modes, and (2) assessing the diffraction efficiency of the HPM for different incident LG orders.

First, we verify that a precise azimuthal $2\pi$ phase change is imparted by the HPM to the wavefront of an incident optical field via holographic reconstruction \cite{Goodman67SPIE, Kim10SPIE}. Any deviation from this $2\pi$ phase change compromises the orthogonality of cascading LG modes \cite{Willner15AOP, Longman17OE}, resulting in the response of the nFP cavity deviating from the target resonance-free behavior. As illustrated in Fig.~\ref{Fig:HPMphase}a, a Gaussian beam LG$_{0}$ is superposed with the converted mode LG$_{1}$ produced by the HPM from LG$_{0}$ (Fig.~\ref{Fig:HPMphase}b), and the phase profile is retrieved from their interference pattern using off-axis holography (Fig.~\ref{Fig:HPMphase}c) \cite{Kim10SPIE}, which shows a phase change of $\approx\!(0.96\pm0.025)\times2\pi$~rad per pass. This provides near-orthogonal modal conversions after each pass through the HPM. This measurement is repeated for the LG$_{1}$ and LG$_{2}$ modes as reported in Supplementary Figure~2, where the orthogonality of the collinearly propagating modes is confirmed.

\begin{figure}[t!]
\centering
\includegraphics[width=8.6cm]{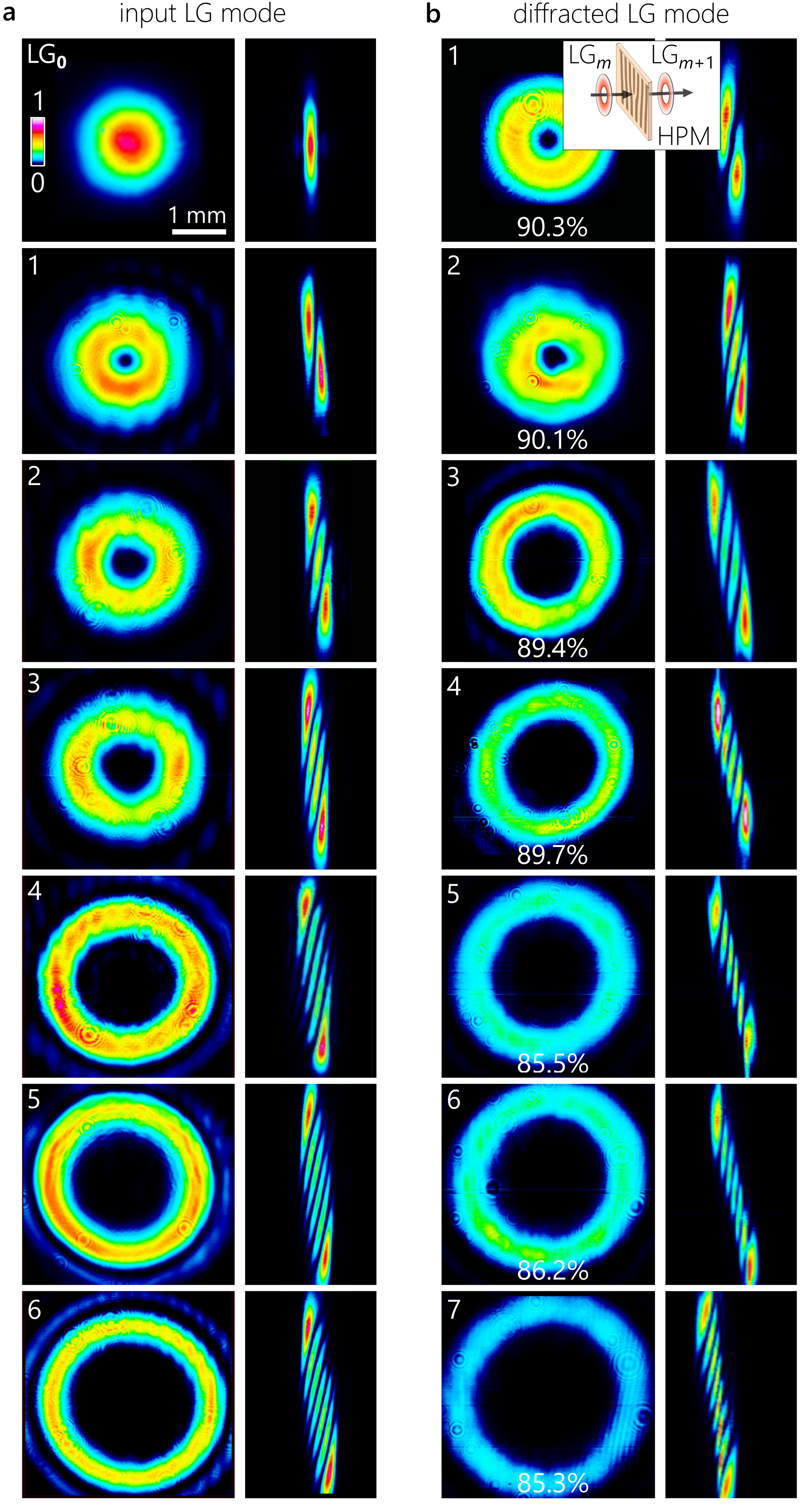}
\caption{\small{\textbf{Modal characterization of the HPM.} (a) The prepared LG modes, from LG$_{0}$ to LG$_{6}$ (Methods). In each row, the left panel is the intensity distribution of the prepared mode, and the right panel verifies the produced modal order via a cylindrical-lens-based technique \cite{Abramochkin91OC, Alperin16OL,Kotlyar17AO}. The number in the top left corner (left panel) indicates the modal order. (b) The diffracted modes generated by the HPM with the input LG modes from (a): LG$_{m}\rightarrow$LG$_{m+1}$. The inset in the top row illustrates the setup for mode conversion with the HPM. The left panel in each row displays the spatial distribution of the diffracted beam after the HPM, and the right panel verifies the produced modal order as in (a). The number in the top left corner (left panel) is the modal order produced. The estimated diffraction efficiency (DE) for the converted order is given below the diffracted field profiles.}}
\label{Fig:HPM}
\end{figure}

Second, we estimate the diffraction efficiency of the HPM across the spectral bandwidth of interest for 7 LG modes. We prepare the modes LG$_{0}$ to LG$_{6}$ (Methods), which are directed to the HPM to undergo diffraction into a higher-order mode LG$_{m}\rightarrow$LG$_{m+1}$ (Fig.~\ref{Fig:HPM}b, inset in top row). The spatial intensity distribution of prepared LG modes (LG$_{0}$ through LG$_{6}$) and the converted modes produced by the HPM (LG$_{1}$ through LG$_{7}$) are shown in Fig.~\ref{Fig:HPM}a and Fig.~\ref{Fig:HPM}b (left panels), respectively. In both cases, the order of the LG field is determined using the astigmatic transform technique \cite{Abramochkin91OC, Alperin16OL,Kotlyar17AO}, which involves analyzing the intensity pattern of the LG beam at the Fourier plane of a cylindrical lens (Fig.~\ref{Fig:HPM}a,b, right panels). In this technique, the OAM number is equal to $n-1$, where $n$ is the number of lobes observed in the Fourier plane of a plano-convex cylindrical lens with a focal length of 150~mm.

The diffraction efficiency is estimated from the ratio of the diffracted power at the optimal Bragg angle of the HPM to the input power, both measured using a calibrated silicon photodetector (Gentec-EO PH100-SI-HA-D0). The HPM used is not provided with anti-reflection coatings, so the measured diffraction efficiency includes a reduction induced by Fresnel reflections. The diffraction efficiency of the HPM for converting the first seven LG modes to their next higher-order counterparts are provided in Fig.~\ref{Fig:HPM}b (at the bottom of the left panels), whose average overall modes exceeds $\approx\!87.5\%$.

\begin{figure*}[t!]
\centering
\includegraphics[width=17.6cm]{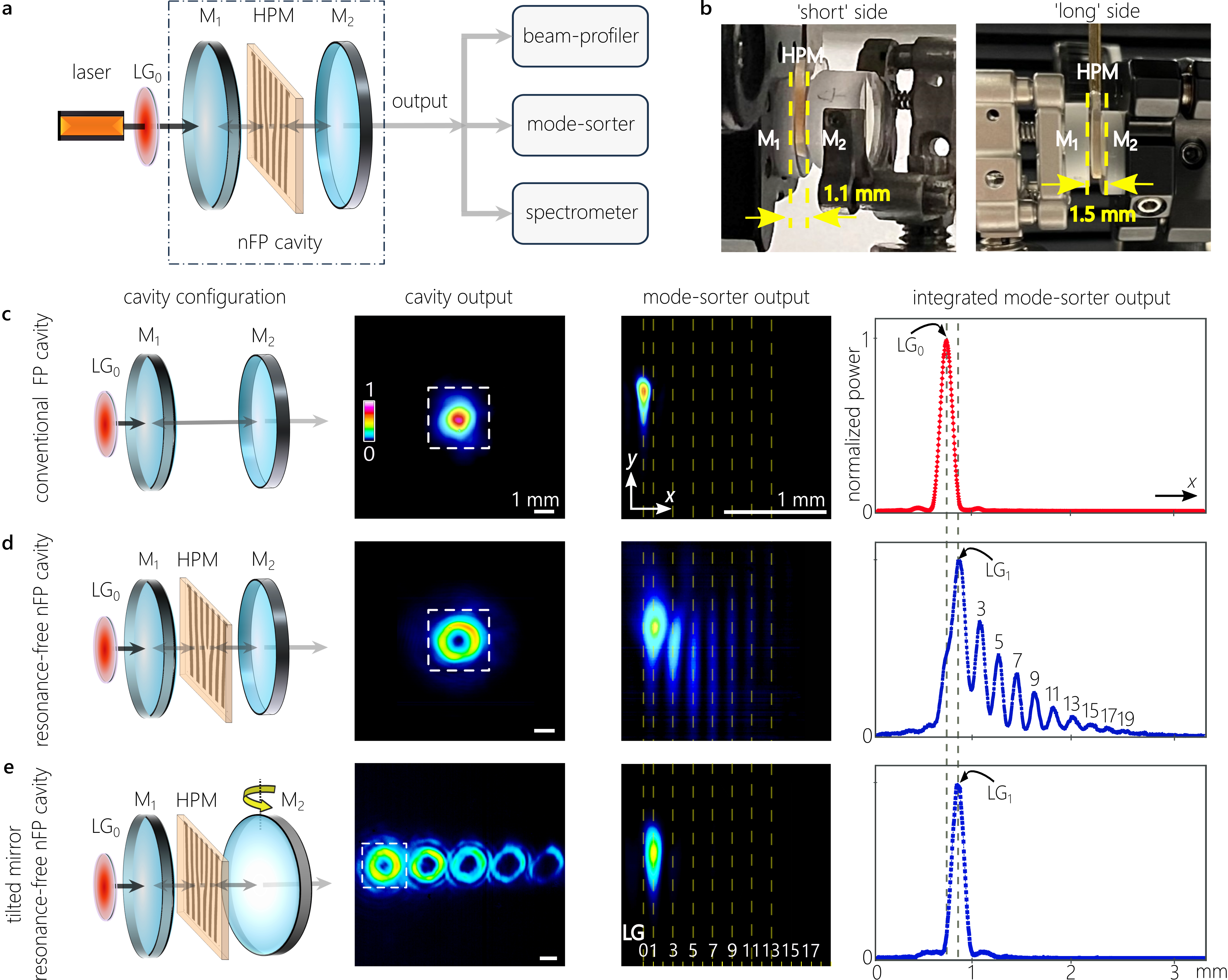}
\caption{\small{\textbf{Modal characterization of resonant FP and resonance-free nFP cavities.} (a) The measurement configuration for the spatial, spectral, and modal characteristics of the cavity output. (b) Photographs of the nFP cavity from both `long' and `short' sides. The mirrors M$_{1}$ and M$_{2}$ are not parallel to accommodate the HPM diffraction angle. The mirrors are depicted in the schematics as parallel for simplicity. (c-e) The spatial intensity distribution of the cavity and mode-sorter outputs for (c) a conventional resonant FP cavity, (d) a resonance-free nFP cavity, and (e) a resonance-free nFP cavity after slightly tilting M$_{2}$ to transversely displace the field each roundtrip. Each row comprises four columns: the first column depicts the cavity configuration; the second displays the spatial distribution of the intensity at the cavity output (the white dashed square indicates the area of the field directed to the LG mode-sorter input); the third displays the output from the LG mode-sorter (the vertical dashed yellow lines correspond to the displaced LG orders); and the fourth column is the integrated power along the $y$-axis of the LG mode-sorter output displayed in the third column.}}
\label{Fig:Cavity}
\end{figure*}

\noindent\textbf{Modal characterization of the nFP.} The base FP cavity consists of two high-reflectivity flat mirrors ($R>94\%$ at a wavelength $\sim1096$~nm), which we illuminate with a continuous-wave Ytterbium fiber laser (YLR-100; IPG Photonics) having a spectral width of 5~nm (full width at $1/e^{2}$ maximum) centered at 1095.6~nm. The randomly polarized laser is rendered vertically polarized via a polarizing beam splitter and a zero-order half-wave plate. The nFP cavity is produced by inserting between the two mirrors an HPM of dimensions $12.5\times12.5\times1$~mm$^{3}$ diffracting at an angle of $6.3^{\circ}$ at the central wavelength, with spectral and angular bandwidths of $\approx32.5$~nm and $\approx4.33$~mrad, respectively (Fig.~\ref{Fig:Cavity}a). The phase pattern constituting the HPM occupies a circular area of diameter $\approx3$~mm in the volume Bragg grating in which it is embedded, which implements the LG mode transformation. Laterally transposing the HPM area to intercept the beam path in the FP cavity allows us to convert the FP cavity into an nFP cavity, and vice versa, by moving the HPM area out of the cavity beam path, all without changing the cavity length. The mirror M$_{2}$ is normal to the diffracted beam from the HPM, and is thus tilted with respect to M$_{1}$. Consequently, the separation between the two mirror edges are different, as shown in the photographs of the nFP cavity from the `long' and `short' sides in Fig.~\ref{Fig:Cavity}b (see also Supplementary Figure~3a).

We depict in Fig.~\ref{Fig:Cavity}a the measurement modalities employed at the cavity exit (see Supplementary Figure~3b-d): (1) spatial beam profiling of the cavity output; (2) analysis of the LG modal content; and (3) measuring the transmitted spectrum. Measurement results for three different cavity configurations are shown in Fig.~\ref{Fig:Cavity}c-e (left column): a conventional FP cavity with the HPM area displaced out of the beam path (Fig.~\ref{Fig:Cavity}c); an nFP cavity incorporating the HPM (Fig.~\ref{Fig:Cavity}d); and an nFP cavity with the out-coupling mirror M$_{2}$ slightly tilted to angularly displace the field after each roundtrip (Fig.~\ref{Fig:Cavity}e). We focus on the transmitted field because the reflected field is dominated by the reflected LG$_{0}$ mode. In all three cases, the cavity is illuminated with a Gaussian beam LG$_{0}$. To record the spatial profile, the cavity is followed by a spherical lens (focal length 150~mm) and the intensity profile is captured by a camera (BladeCam S-BC-XHR) positioned at a distance 120~mm after the lens. 

To analyze the LG modal content of the cavity output, we make use of an LG mode-sorter that separates spatially overlapping LG modes (in a polar coordinate system) into displaced parallel lines (in a Cartesian coordinate system) \cite{Li19OE}. The mode-sorter comprises two axially separated phase plates that implement a log-polar coordinate transformation \cite{Bryngdahl74JOSA,Hossack87JOMO}, after which the azimuthally varying phase distributions of the circularly symmetric LG modes are converted to spatially overlapping, rectangular strips endowed with linearly varying phases, which are spatially delineated using a spherical lens \cite{Berkhout10PRL,Lavery12OE,Li19OE} (Supplementary Section~4). The lateral position of the LG modes after the mode-sorter is $m\lambda\tfrac{f}{d}$, where $m$ is the mode order, $\lambda$ is the wavelength, $f$ is focal length of the lens ($f\!=\!400$~mm here), and $d$ is the length of the transformed beam at the mode-sorter exit \cite{Berkhout10PRL}. This results in a transverse separation of $\approx84$~$\mu$m between adjacent LG-mode orders. A portion of the field at the FP cavity exit (enclosed in a white dashed square in the second column of Fig.~\ref{Fig:Cavity}c-e) is directed to the mode-sorter entrance, and the output from the mode-sorter (third column in Fig.~\ref{Fig:Cavity}c-e) where each mode appears as a vertical strip. In the integrated intensity at the mode-sorter output (fourth column of Fig.~\ref{Fig:Cavity}c-e), each mode appears as a peak. In the case of the conventional FP cavity (Fig.~\ref{Fig:Cavity}c), where the HPM area is displaced out of the cavity, the output field is still Gaussian and matches the input field. After directing the output field to the mode-sorter, only a single vertical strip is visible after the mode-sorter since the cavity transmits only the LG$_{0}$ mode.

By moving the HPM area into the FP cavity, a profound change is observed at the nFP cavity output (Fig.~\ref{Fig:Cavity}d). Rather than a Gaussian mode, we observe a doughnut mode, and upon directing this field to the mode-sorter, we observe a cascade of LG modes reaching to the order $m=19$ (the odd modes from LG$_{1}$ to LG$_{19}$), each diminishing by $\approx14.5\%$ with respect to the preceding order. This drop, and the presence of a remnant LG$_{0}$ mode, is attributed to the finite reflectivity of the cavity mirrors, the efficiency of the HPM, and the deviation from the ideal $2\pi$ phase change per pass. 

Tilting the out-coupling mirror M$_{2}$ in presence of the HPM in the nFP cavity (Fig.~\ref{Fig:Cavity}e) introduces a small angular displacement upon each cavity pass, so that the lens after the cavity spatially displaces the field from each roundtrip, resulting in spatial separation of the transmitted LG modes. This allows each mode to be spatially isolated and directed to the mode-sorter separately. An example is shown in Fig.~\ref{Fig:Cavity}e, where the LG$_{1}$ mode is sent alone to the mode-sorter, which is verified by the appearance of a single line at the mode-sorter output at the expected transverse location.

\begin{figure}[t!]
\centering
\includegraphics[width=8.6cm]{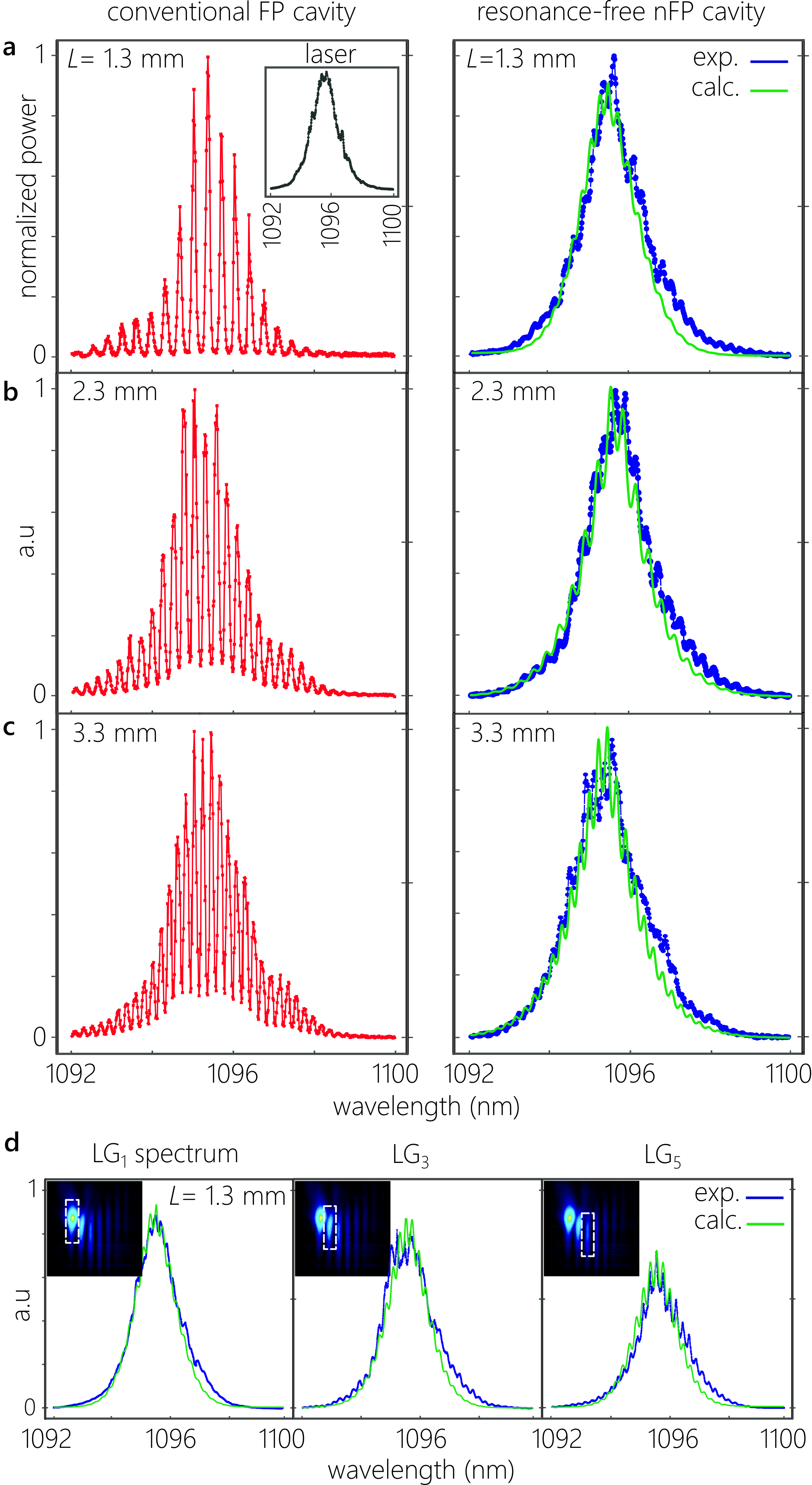}
\caption{\small{\textbf{Spectral transmission for resonant FP and resonance-free nFP cavities.} (a-c) Measured spectra for the conventional resonant FP cavity (left column; corresponding to the configuration in Fig.~\ref{Fig:Cavity}c) and the resonance-free nFP cavity (right column; corresponding to the configuration in Fig.~\ref{Fig:Cavity}d) for different cavity lengths of (a) $L=1.3$~mm, (b) 2.3~mm, and (c) 3.3~mm. Results for other cavity lengths are provided in Supplementary Figure~6. The inset in (a) is the spectrum of the laser source. The green curves in the right column are simulations (see Supplementary Section~6 and Supplementary Figure~7). (d) Measured (exp.) and calculated (calc.) spectra of the first three odd LG orders (LG$_{1}$, LG$_{3}$, and LG$_{5}$) from the nFP cavity (for $L=1.3$~mm), which are spatially separated after the mode-sorter. The inset in each panel is the LG mode-sorter output shows the portion of the field enclosed by a dashed white rectangle (corresponding to a single LG mode) for which the spectrum is measured.}}
\label{Fig:Spectra}
\end{figure}

\noindent\textbf{Spectral characterization.} Finally, we examine the spectrum of the transmitted field. For an ideal HPM, we expect a flat transmission spectral response (limited by the bandwidth of the underlying VBG), so that the output spectrum from the nFP coincides with the input source spectrum. The spectrum of the transmitted field is measured for various cavity lengths $L$ by coupling the cavity output to a multi-mode fiber connected to an optical spectrum analyzer (OSA; Yokogawa AQ6370, spectral resolution $\sim0.02$~nm). We plot in Fig.~\ref{Fig:Spectra} the measured spectra from the conventional FP cavity (left column) and the nFP cavity (right column) for cavity lengths $L\!=\!1.3$, 2.3, and 3.3~mm, corresponding to FSR values of 274, 205, and 150~pm, respectively (measurements for other cavity lengths are provided in Supplementary Fig.~6). The cavity length $L$ is the average of the `long' and `short' sides of the cavity. For example, the $L\!=\!1.3$-mm cavity corresponds to a `short'-side length of 1.1~mm (the edges of the cavity mirrors are nearly in contact) and a `long'-side length of 1.5~mm (Fig.~\ref{Fig:Cavity}b). 

The measured spectra for the conventional FP cavity exhibit a clear resonant response with the expected FSR. In contrast, the transmission spectrum for the nFP exhibits a nearly smooth spectrum that closely matches the input source spectrum (Fig.\ref{Fig:Spectra}a, inset), thereby confirming the resonance-free spectral response from the nFP cavity. A residual resonant response is observed atop the resonance-free spectrum, which is attributed to the imperfections of the HPM: its finite conversion efficiency and deviation from the ideal $2\pi$ phase change per pass. In addition, diffractive spreading with propagation, along with any cavity misalignment result in mode-mixing and cross-talk, and are thus expected to yield a remnant resonant response superposed with the resonance-free response \cite{Vasnetsov05NJP, Xie15Optica, Liu24OE}. The spectral measurements are compared to predictions based on a theoretical model that is described in Supplementary Section~6. 

The spectra plotted in Fig.~\ref{Fig:Spectra}a-c correspond to the total spatial field distributions emerging from the FP and the nFP cavities. In the case of the nFP, this comprises the large number of LG modes seen at the mode-sorter output in Fig.~\ref{Fig:Cavity}d. In addition, we plot in Fig.~\ref{Fig:Spectra}d the spectra for the first three odd LG modes measured separately after being spatially isolated at the mode-sorter output (for $L=1.3$~mm). The measurements are compared to simulations based on the above-described theoretical model. Each mode undergirding the output field shows no spectral resonant structure. Because the HPM imperfections are expected to exacerbate the higher-order LG modes, we note that the resonant spectral components atop the resonance-free response is stronger for LG$_{5}$ compared to LG$_{1}$ (Fig.~\ref{Fig:Spectra}d).

\section*{Discussion}

We have shown that introducing a single intra-cavity phase element (the HPM) into a conventional planar, compact FP cavity renders its spectral response resonance-free for externally incident, broadband optical fields: light is transmitted through the cavity across a broad continuous spectrum rather than at discrete resonant wavelengths. In contrast, almost all previous studies of optical cavities modified to alter the spatial distribution of the cavity field have typically ignored the spectral degree-of-freedom. The compactness of the cavity implemented here reveals unambiguously the change in spectral response, in contrast to previously modified cavities in which the detection of spectral changes is rendered challenging by the extremely small FSR. An exception is the cascaded-mode cavity in \cite{Ginis23NC} in which a phase plate at one of the mirrors in a waveguide-based cavity modifies its FSR by toggling the cavity field between two orthogonal waveguide modes, in contradistinction to the resonance-free response reported here via an intra-cavity phase mask. The key to the new observation reported here is the use of the specially designed thin HPM that increases the mode order after each cavity roundtrip, retains the collinearity and orthogonality of all the generated modes, and minimizes diffraction by virtue of its large active area without increasing the length of the compact cavity structure. The particular modal basis selected was that of LG modes in which the OAM order is increased by the HPM. The HG-like modal basis used previously in on-chip waveguide-based cavities cannot reproduce this effect without incorporating a large number of individual mode-converters \cite{Chang17arxiv}. Our results point towards an interesting consequence of the spatial dimensionality of the modal basis selected. Whereas 1D HG modes in a Cartesian coordinate system (or 2D HG modes that are separable along the two dimensions) cannot undergo unrestricted modal ladder-up with a single element, LG modes (that require both transverse dimensions in a polar coordinate system) can indeed provide this feature. It is an open question which other modal bases can provide a similar resonance-free spectral response. We conjecture that only circularly symmetric modal bases that can carry OAM are suitable for this purpose. Our results thus have important implications for producing on-chip, waveguide-based resonance-free cavities.

The experimental scheme described here provides a compact configuration that is sufficiently flexible to accommodate a host of different modifications, thus offering a springboard for future research in multiple directions. An immediate question to be tackled is the impact of this nFP cavity on prospects for lasing. Adding a gain medium to an nFP cavity stands to tackle the issue of lasing in cavities with OAM-modifying intracavity elements \cite{Sroor20NP,Wen21NP}. Yet to be explored is the incorporation of the polarization degree-of-freedom to the optical field in the cavity, particularly when spin-orbit interactions are relevant \cite{Aiello15NP,Bliokh15NP}. Other possibilities include exploring the possibility of all-solid, monolithic nFP cavities, utilizing metasurfaces in lieu of HPMs to further reduce the cavity length, and on-chip realizations of nFP cavities. Another possibility is the use of spherical cavity mirrors to circumvent diffractive effects encountered in a planar cavity. By varying the diffraction efficiency for different mode orders, or directing a prescribed superposition of modes to the cavity entrance, one can construct complex modal superpositions, which can be particularly useful in quantum communications and sensing. Indeed, the nFP cavity in its implementation here produces an OAM comb, and it would be intriguing to explore the possibility of combining a laser comb (discrete spectral lines) with an OAM comb (discrete OAM modes), and potentially assigning one OAM mode to each spectral line, a field configuration that has shown to yield light springs \cite{Pariente15OL,Zhao20NC}.

A unique aspect of the resonance-free response of the nFP cavity is its insensitivity to perturbations in the cavity length extending here to $\sim350\%$. Therefore, we can harness the cavity field enhancement without the sensitivity to length perturbations. Note that the resonance-free spectral response of the nFP cavity goes hand-in-hand with spatially multimoded cavity operation. In contradistinction, so-called `degenerate cavities' support a spatially multimoded cavity field but retain the conventional discrete resonant spectral response \cite{Slobodkin22Science}.

Finally, we would like to mention the potential of the nFP cavity equipped with an HPM to explore non-Hermitian and topological optical effects. Interestingly, the discrete set of OAM-carrying LG modes supported by our cavity can be viewed as a discrete lattice in a synthetic dimension. Recent studies have shown the possibility of realizing non-Hermitian degeneracies known as exceptional points (EPs) in similar synthetic dimensions \cite{Yang23SA}. In contrast to previous studies, our nFP cavity exhibits a compact geometry thanks to the thin HPM utilized, which allows us to directly probe the spectral characteristics in addition to the synthetic OAM dimension. We envision that our strategy can be utilized to experimentally realize higher-order EPs in the multidimensional parameter space spanned by the OAM-carrying LG modes and the spectral degree-of-freedom. Moreover, the nFP cavity offers the opportunity to explore non-Hermitian topological models in both the passive and active regimes, with potential applications in ultrasensitive cavity-based sensors \cite{Parto25Light} and unconventional lasers \cite{Liu22Light}.


\section*{Methods}

\textbf{Holographic Phase Mask (HPM).} HPMs represent an advanced wavefront modification technology that offers significant advantages over conventional phase masks \cite{Rakuljic92OL, Glebov09JHS, Gerke10NP, SeGall15OE, Divliansky19OE, Mohammadian21JO}. Crucially, because the phase in conventional elements is encoded typically in the optical path length (e.g., through geometric surface variations), they are thus limited to specific wavelengths, and are susceptible to surface contamination \cite{SeGall15OE, Divliansky19OE, Mohammadian21JO}. In contrast, HPMs encode phase information volumetrically within transmissive Bragg gratings (TBGs)  \cite{Rakuljic92OL, Glebov09JHS, Gerke10NP, SeGall15OE, Divliansky19OE, Mohammadian21JO}, which enables HPMs to maintain intricate wavefront modifications across a broad bandwidth by varying the angle of incidence to meet the Bragg condition at each wavelength, thereby providing greater versatility and broadening their range of applications \cite{SeGall15OE, Divliansky19OE, Mohammadian21JO}. 

\textbf{HPM material.} We record the HPM in a photosensitive medium. Here we make use of photo-thermo-refractive (PTR) glass \cite{Efimov99AO, Lumeau17IMR}, which is an ideal medium due to its low absorption in the visible and near-infrared ($\approx10^{-4}$~cm$^{-1}$) and high laser-damage threshold in both pulsed and continuous-wave modalities \cite{Mohammadian21JO, Efimov99AO}, and multiple HPMs can be multiplexed within the same volume of PTR glass with minimal crosstalk \cite{SeGall15OE}. Current PTR-glass technology allows up to 1000~ppm of refractive-index change, which cannot be bleached by laser radiation, while it possesses minimal absorption and scattering from 300~nm in the near ultraviolet to 2000~nm in the near infrared \cite{Glebov09JHS}. These specific properties of PTR glass enable the fabrication of HPMs with near-unity diffraction efficiency, and an extended degree of tolerance to high average- or peak-power laser beams, mechanical shocks, and elevated temperatures.

\textbf{HPM recording.} The HPM is produced by embedding the desired phase information onto a TBG, and holographically recorded into a thick PTR glass \cite{Rakuljic92OL} using a two-beam interference setup, where a phase element is introduced into one of the beams \cite{SeGall15OE}. The resulting interference pattern in the material contains both the grating structure and the desired phase information. As light travels through the HPM, it undergoes diffraction and simultaneously acquires the encoded phase profile, which remains consistent across a wide range of wavelengths when the Bragg condition is met \cite{Glebov09JHS, Gerke10NP, SeGall15OE, Divliansky19OE, Mohammadian21JO}. Only the diffracted beam incurs the embedded phase distribution, while uncoupled light in the transmitted beam observes minimal to no phase modulation. By designing the TBG with a broad acceptance bandwidth, a complete phase reconstruction can be achieved for all operating frequencies within the allowed spectrum \cite{Gerke10NP, SeGall15OE}.

The phase profile of the first Laguerre-Gaussian (LG) mode \cite{Allen92PRA} is recorded in the volume of a PTR glass using a holographic interference setup. LG modes possess orbital angular momentum (OAM), which manifests as a spiral-shaped phase front, and exhibit an intensity profile resembling a donut or ring \cite{Allen92PRA, Yao11AOP}. Moreover, they satisfy the paraxial wave equation in cylindrical coordinates and constitute a complete orthogonal basis set \cite{Allen92PRA}. The holographic recording setup for a typical HPM follows a standard Mach-Zehnder interferometer setup. A volume of PTR glass is exposed to the standing-wave pattern produced by the interference of the signal and reference UV beams in the interferometer. Once the TBG’s Bragg period is determined, the half-angle separation $\theta$rec between the two interfering beams is adjusted to produce a periodic pattern with a matching modulation period. Upon further thermal development, a permanent change in the local refractive-index is induced, forming the volume Bragg grating \cite{Divliansky19OE}. To construct an HPM, a master phase-plate encoding the desired phase distribution is inserted into one of the interferometric arms \cite{Divliansky19OE}. The phase element’s spatial profile, incurred by the signal beam, is transferred to the photosensitive recording medium, superimposing it onto the plane phase-front of a periodic pattern originated from the interference between the signal and reference beams.

\textbf{Preparing LG modes to characterize the HPM.} We prepared the LG modes (from LG$_{0}$ to LG$_{6}$, shown in Fig.~\ref{Fig:HPM}a) used in testing the performance of the HPM and assessing its diffraction efficiency by passing the fundamental Gaussian mode LG$_{0}$ through an appropriate surface fork diffractive grating (FDG). We produced 6 transmission FDGs with 150~$\mu$m period and the proper singularity using electron-beam lithography. The Gaussian laser beam at a central wavelength of 1095.6~nm impinges upon the surface grating, and the first diffraction order beam is selected using a circular aperture.

\vspace{8mm}
\noindent\textbf{Data availability}

\noindent Data supporting the findings of this study are available within the article and its supplementary information files. All raw data generated during the current study are available from the corresponding authors upon request.

\vspace{8mm}
\noindent\textbf{Acknowledgments}

\noindent
This work was supported by the U.S. Office of Naval Research through MURI Awards N00014-20-1-2789 and N00014-20-1-2558.

\vspace{8mm}
\noindent\textbf{Author contributions}

\noindent 
A.F.A. and I.D. conceived the ideas; S.Y. performed the experiments, analyzed the data, and prepared the figures; M.Y. assisted with the experimental design and figure preparation; O.M. fabricated the HPM; E.G.J. and J.K.M. produced the mode sorter; M.P. constructed the theoretical model; A.F.A. wrote the manuscript with input from all authors; I.D. and A.F.A. supervised the research.

\vspace{8mm}
\noindent\textbf{Competing interests}

\noindent
The authors declare no competing interests.

\bibliography{shay}

\end{document}